\documentclass[aps,preprint,showpacs,floatfix,pra]{revtex4}
\usepackage{graphicx}
\usepackage{dcolumn}
\usepackage{bm}
\usepackage{amssymb}
\usepackage{amsmath}
\begin{document}
\draft
\title{
Resonant ratcheting of a Bose-Einstein condensate
}
\author{ L. Morales-Molina$^{1}$ and S. Flach$^{2}$}

\affiliation{$^1$ Department of Physics, National University of Singapore,
117542, Republic of Singapore}
                  
\affiliation{$^2$ Max-Planck-Institut f\"ur Physik Komplexer
Systeme, N\"othnitzer Str. 38, 01187 Dresden, Germany}
\date{\today}
\begin{abstract}
We study the rectification process of interacting quantum particles in a periodic
potential exposed to the action of an external ac driving.
The breaking of spatio-temporal symmetries leads to
directed motion already in the absence of interactions.
A hallmark of quantum ratcheting is the appearance of resonant
enhancement of the current (Europhys. Lett. 79 (2007) 10007 and
Phys. Rev. A 75 (2007) 063424). Here we study the fate of these
resonances within a Gross-Pitaevskii equation which describes a mean field
interaction between many particles. We find, that the resonance
is i) not destroyed by interactions, 
ii) shifting its location with increasing interaction strength.
We trace the Floquet states of the linear
equations into the nonlinear domain, and show that the resonance
gives rise to an instability and thus to the appearance of
new nonlinear Floquet states, whose transport properties differ
strongly as compared to the case of noninteracting particles.

\end{abstract}

\maketitle

\section{ Introduction}

The breaking of space-time symmetries, and their role in the generation
of directed transport in single particle Hamiltonian ratchets, have been extensively
studied in the classical \cite{Flach1,Flach2,Den} and quantum regimes
\cite{Ketz,EPL,PRA,Qu1,Qu2}.  
Experiments with thermal cold atoms loaded on optical lattices \cite{ren1} demonstrated
the fruitfulness and correctness of the theoretical predictions.
Importantly, the latest studies show a resonant enhancement of 
the current in the quantum regime, due to resonances between Floquet states \cite{EPL,PRA}.
Real experiments involve many atoms, and interaction between them may be tuned, but
will always be left at least at some residual nonzero level. Therefore, the impact of
interactions on quantum ratchets has to be addressed.

In this paper, we study, using the mean-field approach, the generation of directed transport
 of interacting quantum particles in a periodic potential under the action of a
 two harmonic driving. With this approach we mimic the motion of cold
 atoms in an optical lattice under the presence of an external force \cite{ren1},
 but at much lower temperatures, when a Bose-Einstein condensate may form \cite{Weitz}.
 To this end, we investigate the continuation of Floquet states of the
 corresponding linear system into the nonlinear domain. 
 We show that a resonant enhancement
 of the current in the nonlinear regime takes place, which
 results from the resonant interaction between
{\em Nonlinear Floquet states}.
  We derive an analytical expression for the evolution of quasienergies in
the nonlinear regime. Finally we show the relation between the transport
properties of nonlinear Floquet states and the asymptotic current
of an initial state with zero momentum.

\section{Model}

Experimental realizations of ratchets with cold atoms 
may tune the temperatures from mK down to $\mu$K, such that
a Bose-Einstein condensate may form due to interactions
between particles \cite{Weitz}.
The corresponding general equation to be studied is then given by the one-dimensional
Gross-Pitaevskii equation (see e.g. \cite{choi})
\begin{equation}
i \hbar \frac{\partial \Psi(\tau)}{\partial \tau} =\left[-\frac{\hbar^2
  }{2M}\frac{\partial^2}{\partial X^2} +V_{0} \cos(2 k_{L} X)- X e(\tau)\right] 
  \Psi+ \frac{4\pi \hbar^2 a_{s}}{ M} | \Psi |^2 \Psi,  \label{shrogauge0}
\end{equation}
where $a_{s}$ is the s-wave scattering length,
$M$ is the atomic mass,
$k_{L}=\pi/d$ is the optical lattice wave number with optical step
$d$, $V_{0}$ is the periodic potential depth,
and  $e(\tau)$ is a periodic driving force. The wave
function is normalized to the total number of atoms in the condensate and we
define $n_{0}$ as the average uniform atomic density \cite{choi,Arimondo}.

 Introducing the dimensionless variables $x=2k_{L} X$, $t=
 \tau/t_s$, $\psi=\Psi/\sqrt{n_{0}}$, and defining $1/\mu=M/4 \hbar k_{L}^2
t_s$; we transform the system (\ref{shrogauge0}) to the dimensionless equation \cite{poletti}

\begin{equation}
i \mu \frac{\partial \psi(t)}{\partial t}
=H_{0} \psi
+g | \psi |^2 \psi,  \label{shrogauge}
\end{equation}
where the dimensionless one-particle Hamiltonian is 
\begin{equation}\label{hamilto0}
 H_{0}=\frac12\hat p^2 +v_{0}\cos(x)-xE(t),
\end{equation}
with  $\hat p=-i\mu \frac{\textstyle \partial}{\textstyle \partial x}$
and rescaled parameters  $v_{0}=\mu^2 M V_{0} /4\hbar^2 k_{L}^2=1$, $E(t)=\mu^2 M
e(t)/8\hbar^2 k_{L}^3$ and $g= \mu^2 C$  with $C=\pi n_{0} a_{s} /k_{L}^2$
\cite{choi}. The dimensionless ac field $E(t+T)=E(t)$.

As in the linear limit \cite{EPL,PRA}, we consider 
$E(t)=E_{1} \cos[\omega (t-t_{0})]+E_{2}\cos[2\omega (t-t_{0})+\theta]$ with $t_{0}$ as initial time.
 
By using the gauge transformation,
$|\psi \rangle \rightarrow \exp[\frac{i}{\mu}x A(t)] |\psi \rangle$, where
 $A(t)=-E_{1} \sin[\omega
(t-t_{0})]/\omega-E_{2} \sin[2\omega (t-t_{0})+\theta]/2 \omega$ is the vector
potential \cite{EPL,PRA}; we
transform the original one-particle Hamiltonian to 
 
\begin{equation}\label{hamilto}
 H_{0}=\frac12[\hat p-A(t)]^2 +\cos(x).
\end{equation}

\subsection{Linear regime}

Consider the Schr\"odinger equation, the linear limit of
Eqs.(\ref{shrogauge}-\ref{hamilto}). Here, we use a tilde to
denote wavefunctions, quasienergies and other relevant
parameters for the linear regime.

It was shown in \cite{EPL,PRA} that, for the appearance of a dc-current in the
quantum regime, two symmetries need to be broken.
These symmetries are defined in the classical limit as follows \cite{Flach1}.
If $E(t)$ is shift symmetric $E(t)=-E(t+T/2)$, then the
Hamiltonian (\ref{hamilto0}) is invariant under the transformation  
 \begin{equation}
 S_{a}: (x, p, t) \rightarrow (-x, -p, t+T/2). \label{eq:Sa}
 \end{equation} 
 Likewise if $E(t)$ possesses the symmetry $E(t)=E(-t)$, then
 (\ref{hamilto0})  is invariant under the transformation
 \begin{equation}
 S_{b}: (x, p, t) \rightarrow (x, -p, -t). \label{eq:Sb}
 \end{equation}
The Hamiltonian Eq.(\ref{hamilto0}) is a periodic function of
time. Then the solutions, 
 $|\tilde{\psi}(t+t_0)\rangle =
U(t,t_{0})|\tilde{\psi}(t_0)\rangle$, can be characterized by the dynamics of
 the
eigenfunctions of $U(T,t_{0})$ which satisfy the Floquet theorem:
%
\begin{equation}\label{eq:floquet}
 |\tilde{\psi}_{\alpha}(t)\rangle=
e^{-i\frac{ \tilde\epsilon_{\alpha}}{ T} t} |\tilde{\phi}_{\alpha}(t)\rangle, \,\,\,\
|\tilde{\phi}_{\alpha}(t+T)\rangle=|\tilde{\phi}_{\alpha}(t)\rangle.
\end{equation}
The quasienergies $\tilde\epsilon_{\alpha}$ $(-\pi < \tilde{\epsilon}_{\alpha} <
\pi)$ and the Floquet eigenstates can be obtained as solutions of the
eigenvalue problem of the Floquet operator
\begin{equation}\label{Operator}
\tilde{U}(T,t_{0})|\tilde{\psi}_{\alpha}(t_{0})\rangle = e^{-i
  \tilde{\epsilon}_{\alpha}}|\tilde{\psi}_{\alpha}(t_{0})\rangle.
\end{equation}
Due to the discrete translational invariance of Eq.(\ref{hamilto}) and Bloch's theorem all Floquet
states are characterized by a quasimomentum $\kappa$ with
$|\tilde \psi_{\alpha}(x+2\pi)\rangle = {\rm e}^{i \hbar \kappa}
|\tilde \psi_{\alpha}(x)\rangle$.

We choose $\kappa=0$ which corresponds to initial states
where atoms equally populate all (or many) wells of the spatial
potential. This allows us to use  periodic boundary conditions
for Eq.(\ref{shrogauge}), with spatial period $L=2 \pi$, so that the
 wave function can be expanded in the plane wave
 eigenbasis of the momentum operator
$\hat{p}$, $|n \rangle=\frac{1}{\sqrt{2 \pi}} e^{i n x}$, viz.

\begin{equation}\label{Eq:wave}
|\tilde \psi(t)\rangle=\sum_{n=-N}^{N} c_{n}(t) |n \rangle.
\end{equation}

Thus, the Floquet operator is obtained by solving
Eqs.(\ref{shrogauge}-\ref{hamilto}) in the linear limit. 
 In the computations we neglect the contribution originating from
 $A(t)^2$, since it only yields a global phase factor. Details of
 the  numerical method are given in \cite{PRA}. 
A study of a related problem with nonzero $\kappa$ has been published in \cite{Ketz},
which shows that the essential features of avoided crossings survive.
\begin{figure}
 \begin{center}
\includegraphics[width=13cm,height=10cm]{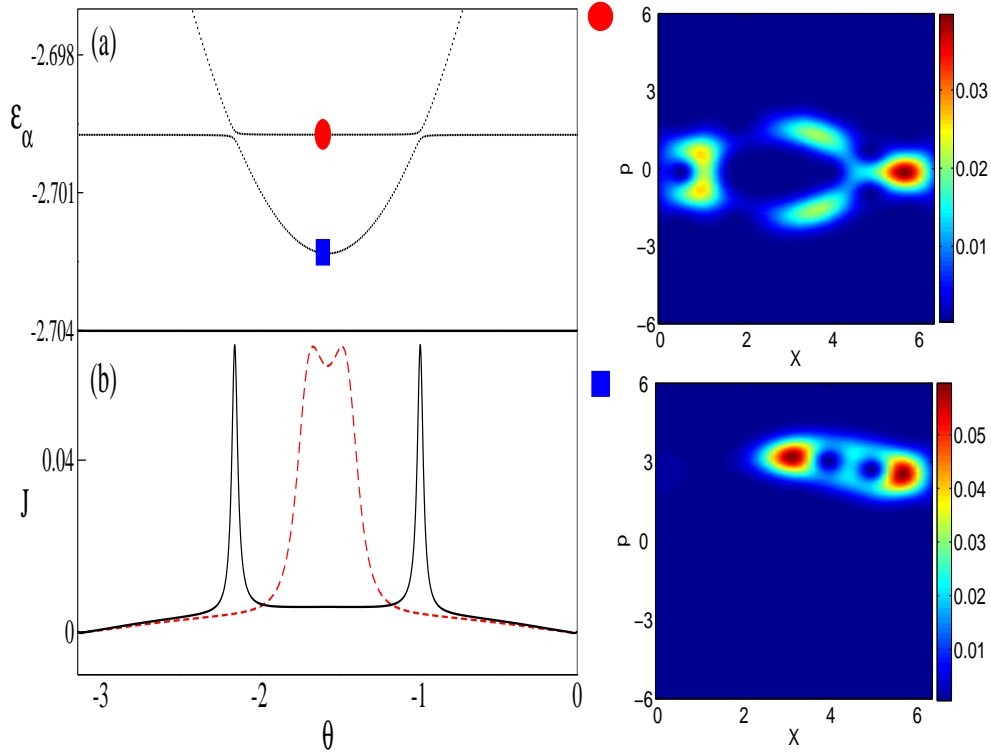}
\caption{Left panel:(a)  Two bands of quasienergies in the linear regime
 whose interaction leads to avoided crossings.
Symbols placed in different bands correspond to states with Husimi functions
depicted in the right panel. Here $E_2=1.2$.
(b) Average current $J$
vs $\theta$ for different amplitude values of the second harmonic
$E_{2}$: $1$ (dashed line) and $1.2$ (solid line). 
Right panel: Husimi functions of 
 two different Floquet states labeled 
 by a circle and square for $t_{0}=0$. We use the Husimi function defined in \cite{PRA}.
Bands and current are plotted for the interval $(-\pi,0)$.
Here the quasienergies are symmetric with respect to a reflection in $\theta$ as deduced from
Eq.(\ref{Sym-ener}), 
while from Eq.(\ref{Sym-current}) it follows that the current is antisymmetric with 
respect to a reflection at the origin.
The other parameters are $E_{1}=3.26$ and $\omega=3$.}\label{Fig:linear}
\end{center}
\end{figure}
Once the Floquet evolution operator is computed, one finds that the
symmetries of the classical equations of motion are
reflected by corresponding symmetries of the Floquet operator.
If the Hamiltonian is invariant under the shift symmetry $S_{a}$
(\ref{eq:Sa}), then the Floquet operator possesses the property
$U(T,t_{0})=U^{\maltese T}\left(T/2,t_{0}\right)U \left(T/2,t_{0}\right)$ \cite{EPL,PRA}.
Here $U^{\maltese}$ performs a transposition along the codiagonal of $U$.
With our driving function $E(t)$, $S_a$ is always violated.
If the Hamiltonian is invariant under
the time reversal symmetry $S_{b}$ (\ref{eq:Sb}), then the Floquet
operator has the property 
$U(T,t_{0})=U(T,t_{0})^{\maltese}$ \cite{EPL,PRA,Graham}.

To estimate the net transport, it is necessary to compute the asymptotic
current. 
It is obtained using the expression $
J(t_0)=\sum_{\alpha} \langle p \rangle_{\alpha}
|C_{\alpha}(t_{0})|^{2}$
 \cite{PRA}, where $\langle p \rangle_{\alpha}$ are the Floquet states
 momenta and $C_{\alpha}(t_{0})$ are the expansion
 coefficients of the initial wave function in the basis of Floquet states.
  Breaking
  the symmetries
  $S_{a}$ (\ref{eq:Sa}) and $S_{b}$ (\ref{eq:Sb}), we 
  desymmetrize the Floquet states, i.e. the Floquet states momenta
 acquire a finite value  $\langle p \rangle_{\alpha}\neq 0$, which
  results in the appearance of a directed transport.

 In general, the current is a function of the initial time $t_{0}$ and the
 relative phase $\theta$, namely $J(t_0,\theta)$. After averaging over the
 initial time it exhibits the property \cite{EPL,PRA}:

\begin{equation}\label{Sym-current}
J(\theta) =-J(\theta+\pi) = -J(-\theta).
\end{equation}  

 We focus the analysis on previous computations obtained for $\mu=0.2$ in
\cite{EPL,PRA}. 

 Fig.\ref{Fig:linear}a shows a section of the quasienergy
spectrum of the Floquet states versus $\theta$ for the interval
$[-\pi,0]$, where a resonance of states, i.e. an avoided crossing, 
 takes place.
  The quasienergies  of the system Eqs.(\ref{shrogauge}-\ref{hamilto}) in the
  linear limit possess the property \cite{EPL,PRA}
\begin{equation}\label{Sym-ener}
\tilde{\epsilon}_{\alpha}(\theta)=\tilde{\epsilon}_{\alpha}(-\theta).
 \end{equation} 
   
 Fig.\ref{Fig:linear}b shows the current dependence on $\theta$. The computation is performed taking the 
 initial state $|0\rangle=1/\sqrt{2\pi}$, which overlaps with states in the chaotic
 layer. With this initial condition we mimic a dilute gas
of atoms which are spread all over the lattice with zero momentum.
  The current has two peaks which are linked  to avoided crossings displayed in Fig.\ref{Fig:linear}a.
 In these particular avoided crossings, states from the chaotic layer
  and transporting states  mix, which leads to a leakage 
 from the chaotic layer to the transporting state, thereby enhancing
 the current.
  
On the other hand, it was shown in \cite{EPL,PRA} that by tuning the amplitude
of the second harmonic of the driving force the peaks becomes broader which
makes it easier resolving it in experiments (see dashed line in Fig.\ref{Fig:linear}b).
\
\subsection{Nonlinear regime}

  In the nonlinear case the analysis of the generation of directed transport in
 the presence of a driving force is 
  much more complicated, due to possible nonintegrability, classical chaos, 
 and mixing \cite{Thommen}. However,
one can take the Floquet states of the linear problem, and continue
 them as periodic orbits into the nonlinear regime.
 Then these nonlinear Floquet states can be analyzed.
 
  While in the linear regime the evolution of the Floquet state
 is determined by Eq.(\ref{Operator}) with a unitary operator $\tilde{U}$,
   in the nonlinear regime the unitarity is lost, and is replaced locally
 by  simplectic maps  \cite{Liu}. 
  Nevertheless, a similar transformation over one period of the ac driving  
  can be defined, viz.
\begin{equation}\label{Osymplectic}
 U\psi_{\alpha}(0)=\exp(-i \epsilon_{\alpha}) \psi_{\alpha}(0),
\end{equation}
 where $U$ is  a nonlinear map of the phase space onto itself, defined
 by integrating a given trajectory over one period of the ac driving \cite{comput}.
 The solutions $\psi_{\alpha}$ constitute generalizations of the linear Floquet
 states (\ref{eq:floquet}).

 To compute the nonlinear Floquet states, we use a numerical  method implemented in 
 computational studies of periodic orbits \cite{comput}. The basics steps are as follows.
 First, we choose a linear Floquet state as an initial seed.
Then taking a small value of the nonlinearity strength, we compute the new solution
using a Newton-Raphson
 iterative procedure, by variying  
 initial seed (see Appendix A for a detailed
 explanation). The procedure involves conservation of the norm and the 
 variation of the quasienergy of the state, which together
 enforce the convergence to the
 desired solution. 
 In each iteration step, the new trial solution is
 integrated over one time period $T$. 
 Once a solution is found, we increase the nonlinearity strength 
 again by a small amount and repeat the same procedure.  
 Thus, we trace the solution into the nonlinear domain.

 Desymmetrization of the Floquet states, due to
  breaking of symmetries, leads to the appearance of directed
  transport in the linear regime with enhancement of transport due to
  resonant Floquet states.
 It is therefore worthwhile to investigate what happens with
  nonlinear Floquet states in the absence of those symmetries,  and trace the fate
  of the abovementioned resonances in the nonlinear regime. 

\subsubsection{Dimer}

 To gain insight into the effect of breaking symmetries on nonlinear Floquet
 states, we utilize a basic model of two coupled BEC states in the presence of an external driving.
  The equations for a driven two sites model can be written as
\begin{eqnarray}\label{Eq:dimer1}
i \mu\,\ \frac{\partial \psi_{1}}{\partial t}=C \psi_{2}+g\,\psi_{1} N _{1} +\psi_{1} f(t),
\\
i \mu\,\ \frac{\partial \psi_{2}}{\partial t}=C \psi_{1}+g\,\psi_{2} N_{2}-\psi_{2} f(t),\label{Eq:dimer2}
\end{eqnarray}
  where $f(t)=f_{1}\sin (\omega t)+f_{2}\sin (2\omega t+\theta) $, $C$ is the
  coupling term, and $N_{1,2}=|\psi_{1,2}|^2$ are the
  populations or number of particles in the sites 1,2.
  The above equations can be also qualitatively  
  viewed as a restriction of the original case (\ref{hamilto})
  to just two basis states with opposite momenta. That leads to the corresponding
  different signs of the last terms in the rhs of the above
  equations.

With $f(t)=0$ 
 the equations above are used to describe, on the mean field
 level, the self-trapping transition of two
 coupled BEC. It was shown in \cite{kladko,smerzi} that
 such a phenomenon occurs when the nonlinearity exceeds a critical value, and the new states
 are characterized by a population imbalance $N_1-N_2$. The
 existence of critical or threshold values have been predicted as well 
 \cite{kladko}. The critical value is 
 determined by the bifurcation point of the stationary solutions. 
 The selftrapped states violate the permutational symmetry of (\ref{Eq:dimer1}),(\ref{Eq:dimer2}),
 which implies that the equations are invariant under permutation of the two indices.
 If $f(t)\neq 0$, the permutational symmetry is broken in general. If however $f(t)$ is symmetric,
 then the dimer equations are invariant under the combined 
 action of permutation {\sl and} time reversal.

Let us consider the linear case $g=0$ first. For $f(t)=0$ the stationary solutions
are the in- and out-of-phase modes $\psi_1(t)=\pm \psi_2(t)=\frac{1}{\sqrt{2}}{\rm e}^{\mp i\frac{C}{\mu}t}$,
which are in fact strictly time-periodic states, with period $2\pi\mu/C$.
Adding the ac drive $f(t)$ they transform into two linear Floquet states. They will
start to become asymmetric, since the presence of $f(t)$ violates permutational invariance,
{\sl but only if $f(t)$ is also violating time-reversal antisymmetry}. 
Note that if $f(t)$ is antisymmetric, the original ac field $E(t)$ is symmetric.
If on the contrary
time-reversal antisymmetry is in place, the linear Floquet states will still be invariant
under the combined action of permutation, time reversal, and complex conjugation.
Thus they do not acquire any population imbalance.
Also the Floquet states will now acquire a nonzero phase shift 
${\rm e}^{-i\nu T}$, where $\nu \approx \frac{C}{\mu}$ in the
limit $f(t) \rightarrow 0$, when iterated over one period of the driving $T$.

In the presence of nonlinearity close to these linear Floquet states there will be again
states, which are 'periodic' in the sense that after one period of the ac driving the
state returns to itself, up to a corresponding phase. Note that both the shape of the
eigenstate, but also the phase (i.e. the analogue of the quasienergy in the
linear case) will smoothly change upon continuation into the nonlinear regime.
The continuation process of such a nonlinear Floquet state is thus encoded by
the linear Floquet state at $g=0$, which is chosen to be continued.

In the case of $f_{1}$, $f_{2}$ being nonzero, $\theta$ becomes a relevant parameter.
For $\theta \neq 0,\pi$ time-reversal antisymmetry is broken, and the nonlinear Floquet
solutions will also loose that symmetry, together with the permutational symmetry (see above).
 
Hereafter, we name by symmetric the case when $\theta=0,\pi$ and nonsymmetric otherwise.
We will also coin linear and nonlinear Floquet states {\sl periodic orbits},
although they are only periodic up to the above mentioned phase shift.
\begin{figure}
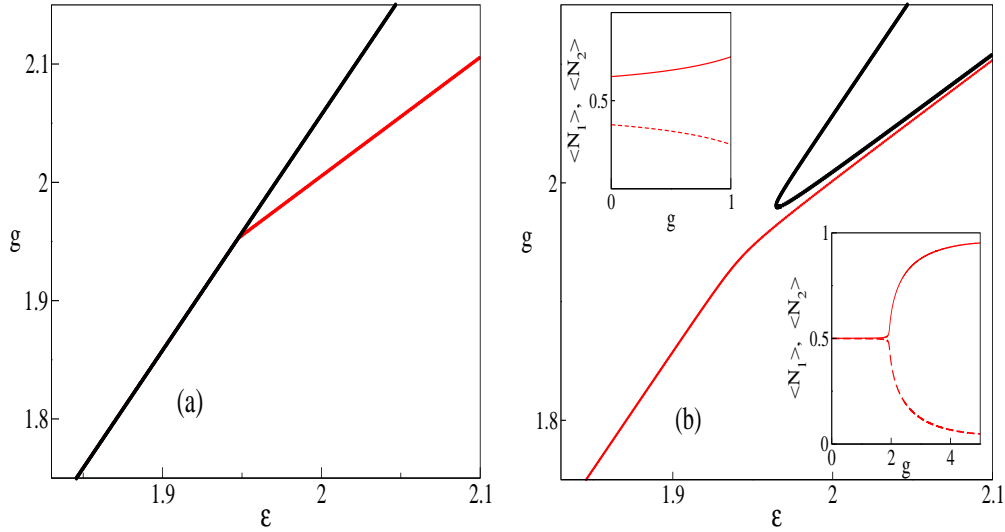

 \begin{center}
\begin{tabular}{lc}
\includegraphics[width=6.43cm,height=7cm]{Fig2a.eps}&
\hspace{0.1cm}
\includegraphics[width=6.43cm,height=7cm]{Fig2b.eps}\vspace{0.4cm}
\end{tabular}\caption{Nonlinearity parameter $g$ 
vs quasienergies for one of the  nonlinear Floquet states. 
(a): $\theta=0$. (b): $\theta=-1.6$. Right-down inset shows the 
  average populations over a period $T$ as a function of the nonlinearity strength $g$. Left-up
  inset is the enlargement around $g=0$ of the right-down inset.
 The  parameters are $\omega=2\pi$, $f_{1}=f_{2}=1$, $C=1$, $\mu=1$.
 }\label{Fig:bif}
\end{center}
\end{figure}

Fig. \ref{Fig:bif} shows the result of continuation of one periodic orbit for the
symmetric and non-symmetric driven dimer, which corresponds to the symmetric
eigenstate of the linear undriven system. The symmetric
 case shows a pitchfork bifurcation with the appearance of two new states at the
 bifurcation point (Fig.\ref{Fig:bif}a).
 These new
 states exhibit a population imbalance similar to the undriven system
 \cite{eilbeck,esser,flach,kladko}, despite the fact that the equations are symmetric.

 This bifurcation is a hallmark of the presence of nonlinearity. Out of a given
 Floquet state several new states are emerging, while for the linear case
 the number of linear Floquet states is fixed by the size of the chosen basis.

 Conversely, if the time-reversal symmetry is broken,  a saddle-node
 bifurcation appears. First of all the state continued from the linear limit,
 already acquires some nonzero population imbalance, since the symmetry
 is broken. 
 Fig.\ref{Fig:bif} shows that the strict continuation
 of that state evolves into a state with a strong population imbalance.
 Two other states - one which is corresponding to a weak imbalance, and another
 which has a strong imbalance as well - emerge through the saddle-node bifurcation.

To conclude this part, we may expect that nonlinearity induces Floquet
states with nonzero population imbalance via bifurcations, or enhances
the already present imbalance (originating from a symmetry breaking) again via bifurcations.
We remind the reader here, that the simple dimer model can be obtained from 
(\ref{hamilto}) by choosing two momentum basis states with different sign,
and replacing the original complicated interaction which is mediated via further basis states,
by a direct interaction term.
A population imbalance thus means a momentum imbalance as well, i.e. a nonzero current.
It may be thus as well expected, that for the full problem, to be treated below,
nonlinearity may enhance directed currents via bifurcations.

\subsubsection{Full lattice}

\begin{figure}
 \begin{center}
\begin{tabular}{lc}
\includegraphics[width=9cm,height=8cm]{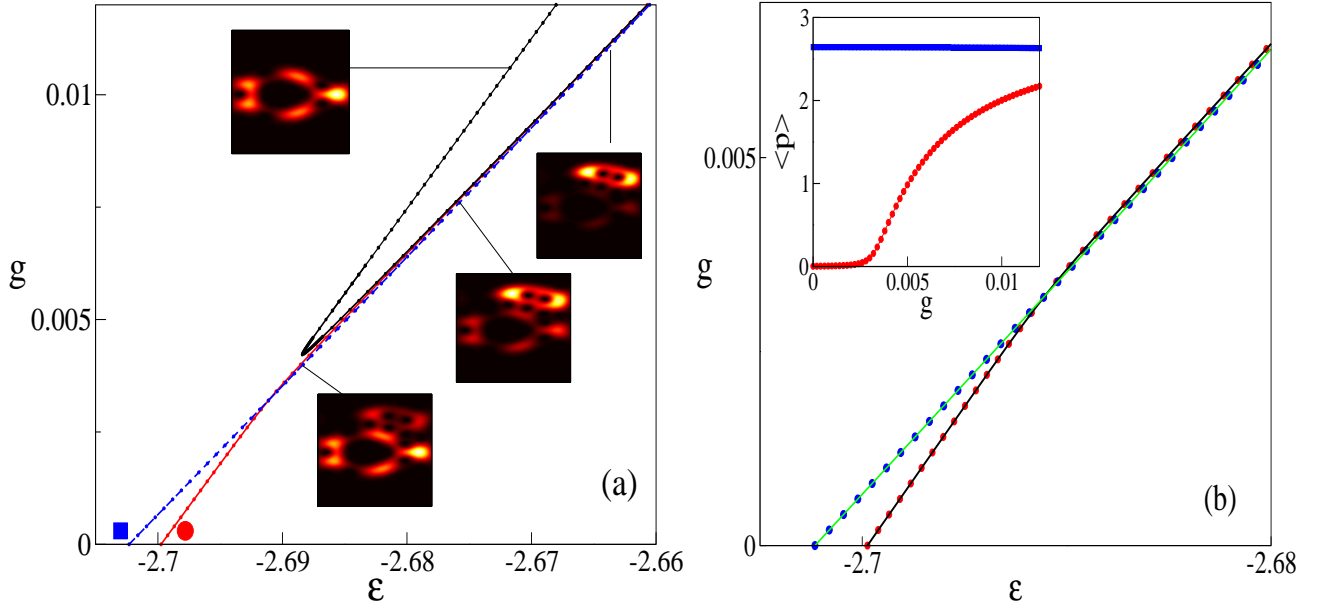}\vspace{0.4cm}
\includegraphics[width=8cm,height=8cm]{Fig3b.eps}
\end{tabular}
\caption{(a): Nonlinearity strength vs quasienergy of periodic solutions.  The
periodic solutions are a continuation from the Floquet states displayed in
Fig.\ref{Fig:linear}. The symbols link to states displayed 
in Fig.\ref{Fig:linear}. 
Insets: Husimi functions ($P$ vs $X$) (cf. Fig.\ref{Fig:linear}). 
The sequence of Husimi functions for the red line is
explained in the text.
(b): Enlargement of the region where the quasienergies of the
continued original states intersect in panel (a). The results from
 the computation of Eq. (\ref{eq:energy-non1}) appear superimposed to the circles with solid lines.
Inset: Mean average momentum of periodic solutions vs nonlinearity strength. blue square: momentum of
  the state depicted by a blue circles-dashed line in (a). red circle:
  momentum  of the 
 state depicted by a red circles-dashed line in (a). The
 parameters are $E_{1}=3.26$, $E_{1}=1.2$, $\theta=-1.6$ and $\omega=3$.
 }\label{Fig:tangent}
\end{center}
\end{figure} 

 The analysis of nonlinear Floquet states in the dimer suggests the  
 formation of states with nonzero mean momentum above a certain threshold value.
 The question then arises, to what extent those features, exhibited by the
 Floquet states in the dimer, are manifested in the full lattice 
  Eqs.(\ref{shrogauge}-\ref{hamilto}). 

  On the other side, the question about the fate of the resonances
  found in the linear limit remains. 
  It is therefore of particular interest to analyse those resonant Floquet
 states, which lead to an enhancement of transport in the linear regime.
 By switching on the nonlinearity in Eq.(\ref{shrogauge}), the system in the
 linear regime gets perturbed and the resonances are shifted to different
 values  of the control parameters. 

 In Fig.\ref{Fig:tangent}, we present the continuations of two 
 Floquet states depicted in Fig.\ref{Fig:linear}, into the nonlinear
 domain.
 We observe a bifurcation of one of the periodic states as we increase the
 nonlinearity strength. 
 Here, as in the dimer, a bifurcation of saddle-node type 
 leads to the formation of  three periodic solutions out of one. 

 Since we want to study the transport of a
 BEC with zero initial momentum, we focus on the continuation 
 of the state lying in the chaotic layer. 
 The continuation process is
indicated by a sequence of Husimi functions in Fig.\ref{Fig:tangent}a,
and can be summarized as follows:
i) before the bifurcation, the
state is located in the chaotic layer; ii) after the bifurcation, the periodic
state transforms into a mixed state due to a resonance with a second transporting  
 state;
 iii) further increase of the nonlinearity strength transforms it into a transporting state.
By contrast, the originally transporting state does not experience significant changes
 as we increase the nonlinearity strength. 

A quantitative measure of this transition process is given by the 
evolution of the average momentum of the periodic states as we change the 
nonlinearity strength [Fig.\ref{Fig:tangent}b].  
The momentum of the state initially located in the chaotic layer goes from
small to high values with a sharp increase at $g\approx 0.003$ (i.e. at the
bifurcation point), whereas
 the momentum for the transporting state remains nearly a constant.

So far, we have analyzed the evolution of two states
 in the nonlinear domain. A clear 
 message from the above results is, that nonlinear Floquet states 
 may drastically change their average momentum at bifurcations,
 which are also visible through sharp changes in the dependence
 of their quasienergies on the nonlinearity parameter.

In order to estimate the shift of the bifurcation point, and to finally
predict the new possible resonance positions with increasing nonlinearity,
we use a perturbation approach to estimate the dependence
$\epsilon=P(\tilde\epsilon,g)$, related to the Aharanov-Anandan
phase \cite{fase} (see also \cite{Liu}).
Assume that nonlinear periodic solutions and Floquet 
states are related through a dynamical
 phase $\lambda(t)$, i. e.,  
 $\phi(t)=\exp[-i \lambda(t)] \tilde\phi(t)$ with $\lambda(T)-\lambda(0)=2k\pi$, $k=0,\pm1,\pm2,..$.   

 Then for $k=0$ we find (see Appendix B for details)
\begin{equation}\label{eq:energy-non0}
\epsilon(\theta,g)= \tilde\epsilon(\theta)+ \frac{g }{\mu}\int_{0}^{T} dt \langle \phi |
|\phi|^2 |\phi \rangle,
\end{equation} 
where  $\langle ... \rangle=\int_{0}^{2\pi}...dx $ is the inner product defined in the 
Hilbert space of Floquet states. 
From Eq.(\ref{eq:energy-non0}) it follows that states with strong nonlinear
interaction have large quasienergy variations. 
That allows to resonantly couple states located deep in the well,
which exhibit strong localization, with other states.
These resonant couplings are beyond the dimer model discussed above. 

Bifurcations of new states correspond to a coalescence of different families of states,
leading to 
 $\phi_{j}(t)=\exp[-i \lambda_{j}(t)] (a_{j}
  \tilde\phi_{1}(t)+b_{j} \tilde\phi_{2}(t))$, where $\tilde\phi_{1}(t)$ and 
$\tilde\phi_{2}(t)$ are
  the corresponding original Floquet states.  
The evolution of the quasienergies for the
  nonlinear periodic states is thus given by 
\begin{equation}\label{eq:energy-non1}
 \epsilon_{j}(\theta,g)=|a_{j}|^2\tilde{\epsilon}_{1}(\theta)+|b_{j}|^2\tilde{\epsilon}_{2}(\theta)+ 
 \frac{g }{\mu}\int_{0}^{T} dt \langle \phi_{j} |
|\phi_{j}|^2 |\phi_{j} \rangle,
\end{equation}

where $a_{j}$ and $b_{j}$ are the corresponding weightings of the linear states in the
nonlinear state expansion (see Appendix B for details).

In Fig.\ref{Fig:tangent}b, we plot the quasienergy values
computed with Eq.(\ref{eq:energy-non1}), using the nonlinear
states continued from the original linear states depicted in
Fig.\ref{Fig:linear}. We find an excellent agreement with the full
numerical results.

For weak nonlinearity, the quasienergies of the states depend linearly on $g$.
A simple perturbation expansion allows to take the wavefunctions of the
original linear states and its respective quasienergies with
Eq.(\ref{eq:energy-non0}). Then from the quasienergy intersection of the two states,
$\epsilon_{1}(\theta,g)=\epsilon_{2}(\theta,g)$, we compute the critical
value of $g$:
\begin{equation}\label{Eq:ecuacionnueva}
g=\frac{\mu [\tilde{\epsilon}_{2}(\theta)-\tilde{\epsilon}_{1}(\theta)]}{\int_{0}^{T}
  \int_{0}^{2\pi}
|\tilde{\phi}_{1}|^4 dx dt-\int_{0}^{T}\int_{0}^{2\pi}|\tilde{\phi}_{2}|^4 dxdt}.
\end{equation} 
Inserting the wave function of the original linear states 
depicted in Fig.\ref{Fig:linear} in Eq.(\ref{Eq:ecuacionnueva}), we obtain $g=0.003008$, which is a good
estimate of the nonlinearity strength at the bifurcation point.

\begin{figure}
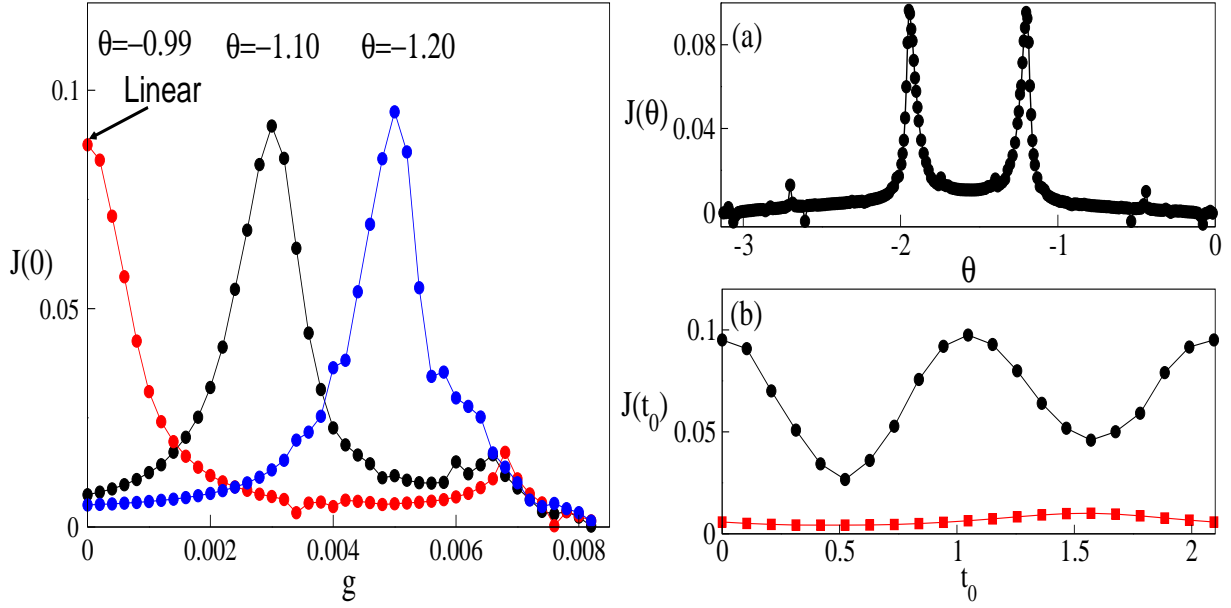

 \begin{center}
\begin{tabular}{lc}
\includegraphics[width=8cm,height=8cm]{Fig4a.eps}\vspace{0.4cm}
\includegraphics[width=8cm,height=8cm]{Fig4b.eps}\end{tabular}
\caption{Left panel: Current dependence upon the  nonlinearity strength $g$
 for $\theta=-1.2, -1.1, -0.99$ with initial time $t_{0}=0$. Right panel:
 (a) Current dependence upon $\theta$ for g=0.005 with $t_{0}=0$. (b) Current as 
 a function of the initial time for
 $\theta=-1.2$ and two different
  nonlinearity strengths. Circles: g=0.005, squares: g=0.001. 
 The other parameters are $E_{1}=3.26$, $E_{1}=1.2$, $\omega=3$.}\label{Fig:init}
\end{center}
\end{figure}

Let us now investigate the evolution of the state $|0\rangle$
 for $\kappa=0$.
In the linear regime, we use the Floquet representation  to derive the
expression for the asymptotic current. In the nonlinear regime it is no longer
possible. Instead, we 
 compute the running average momentum 
 $P=\frac{\textstyle 1}{\textstyle (t -t_{0})} \displaystyle \int_{t_{0}}^{t}\bar p \!\ dt $ with
 $\bar p =\langle \psi|\hat p|\psi \rangle$, over long times, which becomes
 the asymptotic current in
 the limit $t\rightarrow \infty$,
  i.e., $J(t_{0})=\lim_{t \rightarrow \infty} P$. To
 validate the convergence to the asymptotic current we use maximum
 integration times ranging from 100000 till 300000 time periods.

 First we compute the current for $\theta\approx -0.99$. In such a
 case, for $g=0$, there is an avoided crossing (see above discussion), i.e.
 a resonance. Further increase of the
 nonlinearity leads to a decay of the current (Fig.\ref{Fig:init}).
 However, taking $\theta=-1.1, -1.2$,
 we observe a corresponding shift of the resonance peak to larger
 values of the nonlinearity parameter, thus showing a robustness of the resonant current
 enhancement observed 
in the linear limit. 

Fig.\ref{Fig:init} shows the dependence of the current on $\theta$. 
Resonances similar to the linear case are observed, which are shifted
in $\theta$. Thus  the nonlinearity strength
 can also be used as a control parameter for tuning resonances.
 
 It was shown for the linear regime \cite{EPL,PRA} that the
 current depends upon the initial time $t_{0}$. In the nonlinear case, the
 dynamics may exhibit similar behavior to classical chaos \cite{Thommen}, making the 
 analysis more complicated. An essential point here is that, while in the linear regime 
 Floquet states mix in narrow parameter regions via
 avoided crossings,
 in the nonlinear domain mixed states may survive for a fairly large range of
 nonlinearity values. This, along with the fact of having multiple
 bifurcations  with the appearance of new periodic states \cite{berry}, may lead to
 classical-like chaotic behavior predicted in \cite{Thommen}.   
  
  A benchmark for the persistence of directed transport is that 
  the sum of currents for different initial times do not cancel.
  To check that, 
  we compute the momentum evolution for different initial times with the
  system in and out of resonances. In Fig.\ref{Fig:init}, we present the
  current dependence upon the initial time $t_{0}$.
  The curve depicted by filled circles shows a
   similar behavior to the one obtained in the linear
  regime with the system in resonance (cf. Fig.4 in \cite{EPL}).
  It also displays a large positive
  current for all initial times, thus confirming the
 existence of directed motion. On the contrary, the curve with squares shows
 a nearly flat profile with smaller current values, recalling the
 off resonance scenario in the linear limit.

\section{Conclusions}

We have studied the 
rectification of interacting quantum particles in a periodic
potential exposed to the action of an external ac driving, using a mean-field approach. 

We showed that by tuning the nonlinearity in an optical lattice it is
possible to enhance the directed transport of cold atoms. 
A possible experimental way to
achieve it is to vary the scattering length 
by changing the strength of the magnetic field \cite{ketterle}.
These resonances are partly continued from the noninteracting system,
but become saddle node bifurcations of more complicated nonlinear Floquet
states in the nonlinear system under consideration.

We developed analytical estimates of the nonlinearity strength in the resonance,
and showed that the evolution of an initial state with zero momentum
carries all signatures of a ratchet state, i.e. its average momentum is nonzero.
Therefore, ratchets and quantum resonances of (nonlinear) Floquet states
are robust with respect to interactions, and should be observable in real experiments.

Finally, bifurcations of the nonlinear
Floquet states have been observed, which may affect 
the measured currents strongly.

\appendix
\section{Numerical procedure for the computation of nonlinear Floquet
  states}

Take Eq.(\ref{Osymplectic}) 
\begin{equation}\label{Eq:rot}
U\psi_{\alpha}(0)=\exp(-i \epsilon_{\alpha}) \psi_{\alpha}(0),
\end{equation}
where $\psi_{\alpha}$ are periodic solutions in the nonlinear domain and $U$ is a symplectic operator.
\
The periodic states $\psi_{\alpha}(0)$
are decomposed into real and imaginary parts as $\vec{ \Phi }_{\alpha}=\{
{\mbox Real}[ \psi_{\alpha}(0)],
 {\mbox Im}[ \psi_{\alpha}(0)] \}$. Likewise, we define the vector
 $\vec{ \Omega}=\{ {\mbox Real}[\exp(i
  \epsilon_{\alpha}) U \psi_{\alpha}(0)],{\mbox Im}[\exp(i
  \epsilon_{\alpha} ) U \psi_{\alpha}(0)]\} $.
 For convenience we write them as $2N$ 
dimensional vectors $ \vec{ X }_{\alpha}(0)=\vec{ \Phi}_{\alpha}(0) $,
$\vec{X}_{\alpha}=\vec{\Omega}[ \Phi_{\alpha}(0)] $, where $N$ is the dimension
of the Hilbert space \cite{magnus}.
 Hereafter, for simplicity, we drop the index $\alpha$. 

The two vectors $\vec{X}$, $\vec{X}(0)$ are identical for the linear case.
Assuming weak nonlinearity, we compute the vector 
$\vec{X}$ after one integration period, taking as initial seed 
the linear state $\vec{X}(0)$.  This implies that after a full integration
period our final state deviates from the initial state:
\begin{equation}\label{A2}
\vec{G}[\vec{X}(0)]=\vec{X}-\vec{X}(0).
\end{equation}

To correct such a
deviation we use the Newton-Raphson method. Basically, the method updates 
the initial seed $\vec{X}(0)$ after every 
iteration until
each component of $\vec{G}$ is
reduced to a value less than $10^{-9}$.  For
  every integration over the period $T$, we use the split operator method.
 
To successfully accomplish the
 reduction of the vector difference (\ref{A2}), some
constraints should be fulfilled. First, the Floquet states 
 should preserve the norm. We thus add another component $G_{2N+1}=
 \vec{X}\vec{X}-\vec{X}(0)\vec{X}(0)$, and the iteration process is also zeroing this component.

To fully characterize our nonlinear Floquet states, we define the new variable vector 
$\vec{Y}=(\vec{X},\epsilon)$, with
the quasienergy as the $2N+1$ component. Next, we define the new vector function
$\vec{F}(\vec{Y})=[\vec{G}(\vec{Y}),G_{2N+1}(\vec{Y})]$. 
Having $2N+1$ variables and $2N+1$ functions to be zeroed, we now may apply 
standard Newton-Raphson methods.

\section{Quasienergy dependence on the nonlinearity strength}
 We define an expression similar to Eq.(\ref{eq:floquet}), but for the
nonlinear case, viz.

\begin{equation}\label{eq:floquet3}
 |\psi(t)\rangle=
e^{-i\frac{  \epsilon}{ T} t} |\phi(t)\rangle, \,\,\,\
|\phi(t+T)\rangle=|\phi(t)\rangle.
\end{equation}
Here, we have dropped the index $\alpha$.

Then, inserting Eq.(\ref{eq:floquet3}) into Eq.(\ref{shrogauge}), one finds 
\begin{equation}\label{eq:sho1}
i\mu \frac{\partial }{\partial t} |\phi \rangle=\mathcal{H} |\phi\rangle,
\end{equation}
where $\mathcal{H}=H_{0}- \frac{\textstyle \mu }{\textstyle T}\epsilon+ g
|\phi|^2 $. Similarly, using Eq. (\ref{eq:floquet}) we get  
for the linear case 

\begin{equation}\label{eq:sho2}
i\mu \frac{\partial}{\partial t} |\tilde\phi \rangle =\tilde{\mathcal{H}} |\tilde\phi\rangle,
\end{equation}
where  $\tilde{\mathcal{H}}=H_{0}- \frac{\textstyle \mu
 }{\textstyle T} \tilde{\epsilon} $.

Now define $\phi(t)=\exp[-i \lambda(t)] \tilde\phi(t)$ such that
$\lambda(T)-\lambda(0)=2 k \pi$ with $k=0,\pm1,\pm2..$. This assumes that after projecting the nonlinear
state onto the linear space, both vectors, the projected and the
Floquet states in the linear space, have a phase difference
$\lambda(t)$ \cite{fase}. For simplicity we take
in the following $k=0$.

 Using then $\tilde\phi(t)$ and Eq.(\ref{eq:sho2}), we get
\begin{equation}
-\frac{\partial \lambda(t)}{\partial t}= \frac{1}{\mu} \langle \tilde\phi(t)|\tilde{\mathcal{H}}|
 \tilde\phi(t)\rangle- \langle \phi(t)|i
 \frac{\partial}{\partial t}| \phi(t)\rangle.
\end{equation}

Using Eq. (\ref{eq:sho1}) and integrating over a period, we find
\begin{eqnarray}\label{B5}
0=\frac{1}{\mu} \int_{0}^{T}dt [ \langle
\tilde\phi(t)|H_{0}| \tilde\phi(t)\rangle -\langle
\phi(t)|H_{0}| \phi(t)\rangle]
\nonumber \\  - \tilde\epsilon + \epsilon -
\frac{g }{\mu}\int_{0}^{T} dt \langle \phi | |\phi|^2 |\phi \rangle.
\end{eqnarray}

 The first term of the r.h.s. of equation (\ref{B5}) 
is zero after integrating over one period. Then we obtain       

\begin{equation}\label{eq:energy-non}
\epsilon= \tilde\epsilon+ \frac{g }{\mu}\int_{0}^{T} dt \langle \phi |
|\phi|^2 |\phi \rangle.
\end{equation}

As follows from Eq.(\ref{Eq:rot}),
 $\epsilon$ denotes the phase accumulated after a completion of one period
 $T$.
To understand the physical meaning of Eq. (\ref{eq:energy-non}), we 
rewrite it as a function of the real quasienergy $\mathcal{\tilde E}$  (see
\cite{griffoni2} for definition). It relates to our $\tilde{\epsilon}$ as
$\mathcal{\tilde E}=\tilde{\epsilon} \mu/T $, where
$-\mu\,\ \omega /2<\mathcal{\tilde E} < \mu\,\ \omega /2$. Then we obtain

\begin{equation}
 \mathcal{E}=\tilde{\mathcal{E}}+ \frac{g }{T}\int_{0}^{T} dt \langle \phi |
|\phi|^2 |\phi \rangle=\tilde{\mathcal{E}}+\frac{1 }{T}\int_{0}^{T} dt\,\
g \int_{0}^{2\pi} dx \ |\phi|^4.
\end{equation}

This expression tells us that the energy of our nonlinear
Floquet state is the sum of the corresponding linear and nonlinear contributions. The second
term is the time average of the state energy due to the nonlinear interaction.

So far, we have considered the energy evolution of single states without
  perturbation. However,
  single states may ``coalesce'' leading to
  bifurcations. We take
  the simplest case with
   two resonant states. We project vectors on a basis 
   that results from the superposition of the continued original eigenstates, whose
   continuation lead to resonances in the nonlinear domain. That is, 
  $\phi_{j}(t)=\exp[-i \lambda_{j}(t)] (a_{j}
  \tilde\phi_{1}(t)+b_{j} \tilde\phi_{2}(t))$ with $j=1,2$; where $\tilde\phi_{1}(t)$ and
  $\tilde\phi_{2}(t)$ are the original Floquet states. 

By doing similar operations as above we get
\begin{equation}
 \mathcal{E}_{j}=|a_{j}|^2\tilde{\mathcal{E}}_{1}+|b_{j}|^2\tilde{\mathcal{E}}_{2}+ 
 \frac{g }{T}\int_{0}^{T} dt \langle \phi_{j} |
|\phi_{j}|^2 |\phi_{j} \rangle.
\end{equation}

where $|a_{j}|^2=|\langle \tilde\phi_{1}| \phi_{j} \rangle |^2$, and $|b_{j}|^2=1-|a_{j}|^2=|\langle 
\tilde\phi_{2}| \phi_{j} \rangle |^2$.

\end{document}